\documentclass[prl,preprint]{revtex4-1}
 \usepackage{epsfig}
\usepackage{multirow}
\usepackage{natbib}
 
\begin{document}

\title{The Stacking in Bulk and Bilayer Hexagonal Boron Nitride}
\author{Gabriel Constantinescu}
\affiliation{
 School of Engineering and Science, Jacobs University Bremen, 
 Campus Ring 1, 28759 Bremen, Germany}
\author{Agnieszka Kuc}
\email{a.kuc@jacobs-university.de}
\affiliation{
 School of Engineering and Science, Jacobs University Bremen, 
 Campus Ring 1, 28759 Bremen, Germany}
\author{Thomas Heine}
\email{t.heine@jacobs-university.de}
\affiliation{
 School of Engineering and Science, Jacobs University Bremen, 
 Campus Ring 1, 28759 Bremen, Germany}

\date{}

\begin{abstract}

The stacking orders in layered hexagonal boron nitride bulk and bilayers are studied using high-level $ab~initio$ theory (local second-order M\o ller-Plesset perturbation theory, LMP2). Our results show that both electrostatic and London dispersion interactions are responsible for interlayer distance and stacking order, with AA' being the most stable one. The minimum energy sliding path includes only the AA' high-symmetry stacking, and the energy barrier is 3.4 meV per atom for the bilayer. State-of-the-art Density-functionals with and without London dispersion correction fail to correctly describe the interlayer energies with the exception of PBEsol that agrees very well with our LMP2 results and experiment.
\end{abstract}
\maketitle

Layered materials, characterized by weak interlayer forces, result from anisotropy of the bonding energy in different crystal lattice directions.
While the strong intralayer interactions have often covalent character, the much weaker interlayer interactions are dominated by London dispersion forces (van-der-Waals), electrostatic interactions (Coulomb), or a resultant balance between the two.
Many two-dimensional (2D) systems exist in the bulk as layered form.
Among them, the most prominent ones are graphene, hexagonal boron nitride ($h$-BN), transition-metal dichalcogenides (TMDs, e.g.\ MoS$_2$) and oxides (TMOs, e.g.\ titania).
Single layers of 2D materials can be produced by micromechanical cleavage,\cite{Novoselov2005, Pacile2008} liquid exfoliation\cite{Coleman2011} or chemical vapor deposition.\cite{Song2010, Liu2011}
Such confinement of the 3D systems into the 2D forms may result in distinct properties of the latter.\cite{Novoselov2005, Kuc2009}
For example, the parabolic dispersion relation in graphite's band structure changes to a linear band behavior in graphene and can be described by massless Dirac fermions.\cite{Novoselov2005}

Layered materials are predominantly formed in hexagonal symmetries including different stacking orders of the hexagonal layers.
Stacking faults are common as very often the energy barrier between different stackings is small enough and polytypism can occur due to the sliding of layers with respect to
each other.
This phenomenon is widely applied in solid lubricants, such as $h$-BN\cite{Cho2013} or WS$_2$ nanostructures.\cite{ws2pnas}
Polytypism and stacking faults may considerably influence physical properties of layered materials.

The determination of stacking orders is not straight forward, in particular for structurally more involved systems.
For example, first reports on the 2D structures of covalent-organic frameworks (COFs) claim AA and AB stacking orders.\cite{Cote2005, Cote2007}
We have shown that those stackings are energetically less stable compared to inclined and serrated forms, resulting from the Coulomb repulsion between neighboring layers.\cite{Lukose2011}
Serrated and inclined stacking of 2D COFs is now commonly accepted structure in the literature.\cite{Colson2011, Biswal2013}

\begin{figure}[h!]
\begin{center}
\includegraphics[scale=0.33,clip]{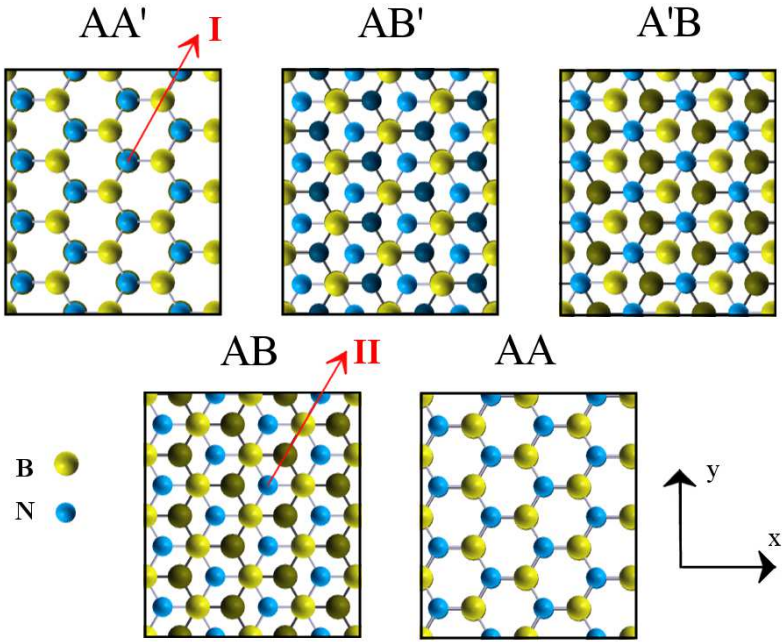}
\caption{\label{fig:1}(Online color) High-symmetry $h$-BN stackings (with depth cueing). The arrows indicate the imposed sliding directions of the alternate layers.}
\end{center}
\end{figure}

Hexagonal boron nitride ($h$-BN) is a layered material, isostructural to graphite except for the stacking order.
Graphite is well-known to crystallize in the AB (staggered) phase, while $h$-BN favors  AA' (eclipsed) stacking.\cite{Pease1950}
In total, there are five high-symmetry stacking orders proposed and investigated for $h$-BN (see Figure~\ref{fig:1}).
These are AA' (eclipsed with B over N), AB' (staggered with B over B), A'B (staggered with N over N), AA (eclipsed with N over N and B over B) and AB (staggered with B over N).
They can be transformed into each other by translational sliding of one of the basal planes in the unit cell: going from the AA' ground state through the AB' to the A'B stacking fault, or by rotation of the basal plane around the $c$ axis and the subsequent sliding mode: going from the AB to the AA stacking fault (see Fig.~\ref{fig:1}).

Recent experimental work of Warner et al.\cite{Warner2010} on the topography of $h$-BN sheets, produced by chemical exfoliation, has shown that besides the AA' ordering, AB stacking is possible and observed for the bilayer regions of this material.
This suggests that $h$-BN might exist in different polytypes depending on the number of layers stacked together.

In parallel to the experiments, several theoretical studies on bulk and a few on bilayer $h$-BN have been performed employing density functional (DFT) or Hartree-Fock (HF) theory.\cite{Liu2003, Ooi2006, Koskilinna2006, Marom2010, Ribeiro2011, Yin2011, Hod2012}
The most common approximation used for the DFT exchange-correlation functional was (semi-)local (LDA) or the generalized gradient approximation (GGA), which is known to fail to describe weak London dispersion interactions correctly.\cite{Lein1999, Klimes2012}
This failure of present density-functionals can be cured efficiently by adding an empirical force-field-like correction on top of the DFT energies (DFT-D) (see for example Ref.~\citenum{Klimes2012} and references therein). 
Applied to $h$-BN, the results of many LDA and GGA calculations predict AA' and AB stackings to be equally stable, both for the bulk and bilayer forms.\cite{Liu2003, Ooi2006, Ribeiro2011, Yin2011}
On the other hand, based on DFT-D calculations, Marom et al.\cite{Marom2010} claimed that the $h$-BN bilayer prefers an AB$_1$ (AB') stacking, a geometry having boron atom over boron atom in the adjacent layer (cf.\ Figure~\ref{fig:1}).
These findings are in contrast with the well-known AA' experimental structure of bulk $h$-BN,\cite{Pease1950} oppose intuition due to the repulsive Coulomb forces between equally charged borons, and disagree with recent DFT calculations.\cite{Ribeiro2011}

In order to resolve this contradiction between various quantum approaches and the experimental results, we have employed local second-order M\o ller-Plesset perturbation theory (LMP2) for solid state\cite{Pisani2008} to study the layer stacking and the lowest-energy sliding path in bulk and bilayer $h$-BN.
Our results show that for both cases the most stable stacking order is AA', followed by  the AB stacking fault.
We further show that DFT-D is able to predict the correct order of the least stable structures, but cannot accurately resolve the energy differences between the most stable stacking forms (see Supporting Information (SI)), except for the PBEsol\cite{PBEsol} functional.

We have employed local-MP2 (LMP2) calculations for the solid state, as implemented in CRYSCOR,\cite{Pisani2008} on the basis of the Hartree-Fock (HF) Bloch orbitals.
The main advantage of the LMP2 scheme is that it treats the system purely quantum mechanically, thus avoiding the empirical nature of force-field-like London dispersion correction schemes.
LMP2 provides (at the moment) probably the most accurate intrinsic description of the London dispersion interaction for periodic systems.
However, recently, the random-phase approximation (RPA) has been applied to a wide range of solids, showing good performance for non-covalent interactions as well.\cite{Bjorkman2012}
Moreover, the improved versions of the Tkatchenko and Scheffler method, in which electrodynamic response effects are included via solving the self-consistent screening equation of electrodynamics and many-body effects, offer a first-principles treatment of the London Dispersion interactions.\cite{Bucko2013}
For comparison, we have performed a series of DFT calculations using various density functionals (PBE0\cite{PBE0, PBE0a}, B3LYP\cite{B3, LYP}, PBE\cite{PBE}, BLYP\cite{B, LYP}, and PBEsol\cite{PBEsol}), with and without  dispersion correction according to the approach of Grimme,\cite{Grimme2006a} as implemented in Crystal09.\cite{CRYSTAL09}
Periodic boundary conditions are employed using shrinking factor set to 12, which corresponds to 133 $k$ points in the irreducible Brillouin zone as proposed by Pack and Monkhorst.\cite{Monkhorst1976}
For the perturbation theory, we have employed excitation domains consisting of six atoms.
We have employed three Gaussian-type basis sets: 6-21G* (denoted as BS1), a rather exhaustive triple-zeta basis set optimized for $h$-BN (denoted as BS2), and BS2 with additional diffuse functions (denoted as BS3).\cite{Halo2011, Halo2011a}
The calculations were corrected for basis set superposition errors (BSSE) using counterpoise method.\cite{Boys1970}
The two larger basis sets (BS2 and BS3) show almost negligible BSSE and LMP2/BS3 gives an energy minimum at the bulk interlayer distance ($c$ = 3.34 \AA) close to the experimental data, while the BS2 shifts $c$ to much higher values (see Figures S1 and S2 in SI).
This shows that additional diffuse functions are essential for the proper description of the electron correlation.
Hereafter, we solely use the results obtained with the BS3 basis set after the BSSE correction, unless stated otherwise.

We have optimized the basal plane of $h$-BN at the HF level and the interlayer distance of the AA' stacking in the bulk and bilayer forms using the LMP2 method.
The lattice parameters and the interlayer distances obtained at the LMP2 and DFT levels are summarized in Table S1 in the SI.
We have obtained the same lattice constant $a$ = 2.49 \AA\  for bulk and bilayer, and also the interlayer distance $c$ = 3.34 \AA\ coincides.
The in-plane lattice $a$, dominated by the covalent interactions, was very well described already at the HF level and agrees perfectly with the experimental value of 2.50 \AA.\cite{Pease1950}
While $c$ exactly matches the experiment for the bulk form (3.33 \AA\cite{Pease1950}), the bilayer interlayer distance is within the experimental error range (3.25$\pm$0.10\cite{Warner2010}).
The $c$ values slightly differ if we ignore the BSSE correction (3.27 \AA\ and 3.31 \AA\ for the bulk and bilayer, respectively) and they are still in a very close agreement with the experimental values.

\begin{figure}[h!]
\begin{center}
\includegraphics[scale=0.2,clip]{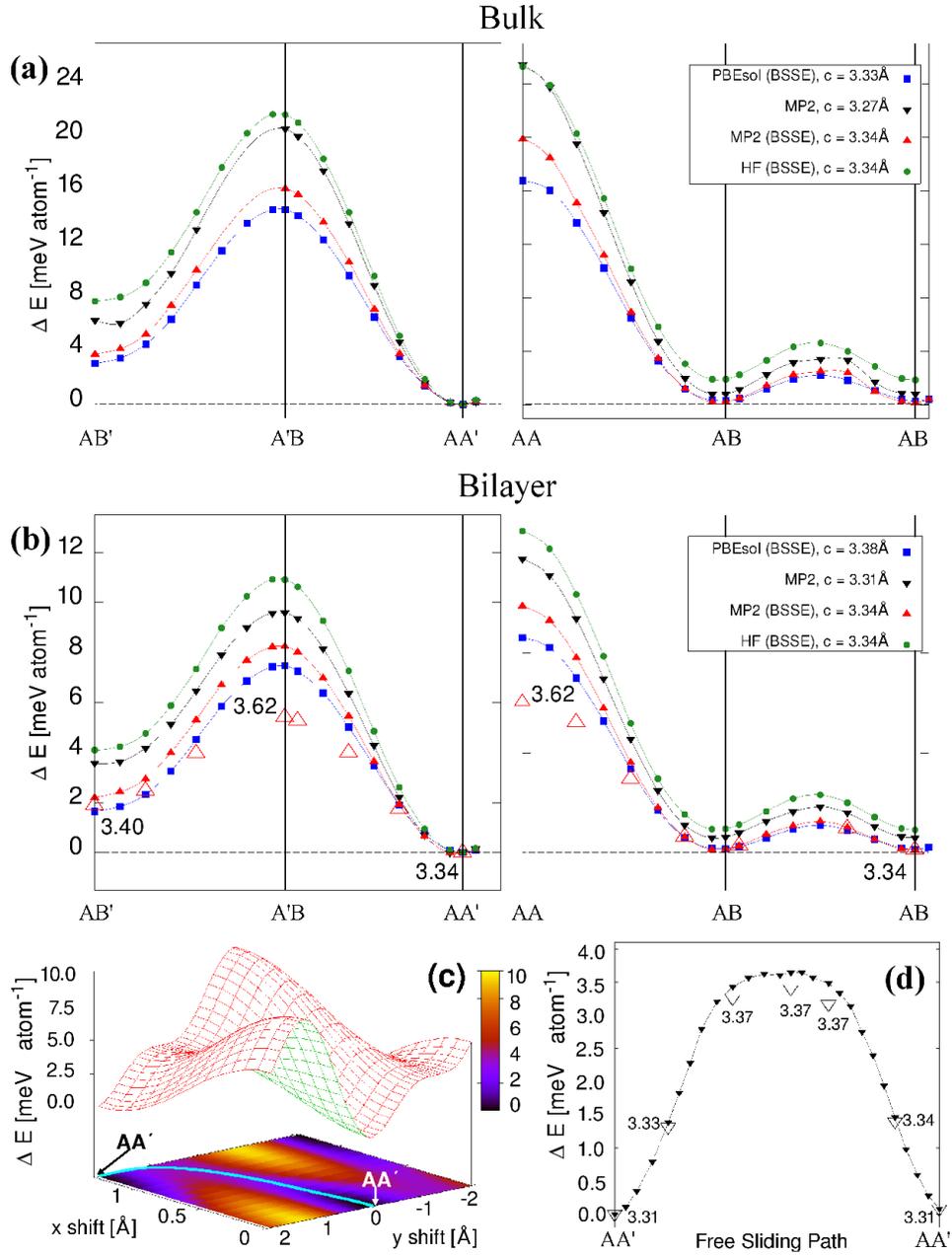}
\caption{\label{fig:2} (Online color) Calculated LMP2 stacking energies of the bulk (a) and bilayer (b) $h$-BN. (c) The energy surface and (d) the lowest-energy sliding path (calculated without the BSSE correction) for the $h$-BN bilayer material. All numbers are related to the equilibrium AA' structure. The energies obtained after $c$ reoptimization are marked with empty symbols, and corresponding interlayer distances (in \AA) are given next to the data point.}
\end{center}
\end{figure}

First, we have investigated the high-symmetry sliding path as shown in Figure~\ref{fig:1}, keeping $c$ fixed.
Thus, we obtained all high-symmetry stacking faults for the bilayer and bulk.
The relative energies (Table~\ref{tab:1} and Figures~\ref{fig:2} a, b) are calculated with respect to the equilibrium structure, that is, they refer to the lowest-energy stacking order AA'.
Corresponding DFT-D calculations are given in the SI.

\begin{table}
\centering
\caption{\label{tab:1} LMP2 sliding energies of the high-symmetry stacking faults in meV per atom for different interlayer distances (in \AA). All the numbers are given with respect to the equilibrium AA' stacking order. The energies are given with and without BSSE.}
\begin{tabular}{c|c|c|cccc}
\hline
 \multirow{2}{*}{\textbf{~System~}} & \textbf{BSSE} & \multirow{2}{*}{ \textit{\textbf{~c~}}}  &  \multirow{2}{*}{\textbf{~AA~}}  &  \multirow{2}{*}{\textbf{~A'B~}}  &  \multirow{2}{*}{\textbf{~AB'~}} &  \multirow{2}{*}{\textbf{~AB~}} \\
& 
\textbf{
corrected
} &&&&\\
\hline
\multicolumn{7}{c}{\textbf{LMP2}} \\
\hline
\multirow{2}{*}{\textbf{bulk}}     & n & ~3.27~ & ~25.39~ & 20.57 & 6.30 & 0.81 \\ 
                                                     & y & ~3.34~ & ~19.89~ & 16.14 & 3.76 & 0.44  \\
&&&&&\\                                                
\multirow{2}{*}{\textbf{bilayer}} &n & ~3.31~ & ~11.75~ & 9.58 & 3.60 & 0.61 \\ 
                                                     & y & ~3.34~ & ~9.86~    & 8.25 & 2.21 & 0.12  \\
\hline
\multicolumn{7}{c}{\textbf{PBEsol}} \\
\hline
\textbf{bulk} & y & ~3.33~&~16.76~&~14.57~& 3.07 & 0.34 \\
\textbf{bilayer} & y &~3.38~&~8.61~&~7.47~& 1.65 & 0.15 \\
\hline                                                
\multicolumn{7}{c}{\textbf{HF}} \\
\hline
\textbf{bulk}      & y & ~3.34~ & ~25.27~ & 21.64 & 7.71 & 1.87  \\ 
\textbf{bilayer} & y & ~3.34~ & ~12.87~ & 10.91 & 4.10 & 0.93  \\                                                
\hline
\end{tabular}
\end{table}

The LMP2 relative energies indicate that the lowest-energy stacking is AA', closely followed by the AB form, which is only by 0.4 and 0.1 meV per atom less stable than the equilibrium for bulk and bilayer, respectively. 
In the case of the bilayer, the small energy difference between the AA' and AB stackings explains the transmission electron microscopy (TEM) results of Warner et al.\cite{Warner2010}, who have observed islands of AB stacking together with larger areas of AA' ordering on the $h$-BN surfaces after activation by the TEM electron beam. In the same work, it has been
reported that AB stacking is indeed possible for  bilayers and for outer layers in the bulk samples.
The AB' stacking fault, reported  to be the most stable for the bilayer by Marom et al.,\cite{Marom2010} is by at least 3.8 meV per atom for the bulk and 2.2 meV per atom for the bilayer less stable than the AA' and AB stackings.

The other high-symmetry stackings, A'B and AA, are significantly destabilized, by more than 8 meV per atom (bilayer) and 16 meV per atom (bulk). 
We account the low stability of these stackings to the Coulomb repulsion, manifested in the eclipsed stacking of N (A'B) or of all atoms in the case of AA.
This is supported by the very similar performance of the HF stacking energy as function of the sliding path, that runs parallel to that of LMP2.
As the Bloch waves, and thus the electronic density, are identical in HF and LMP2, the electrostatic interlayer interactions are equivalent between both methods, and the difference is correlation energy, essentially London dispersion.
Thus, the electrostatic contribution is dominating the sliding energy as already reported by Marom et al.\cite{Marom2010}

We have optimized the interlayer distance $c$ for selected high-symmetry points in the bilayer system.
At the high-energy stacking faults A'B and AA, the interlayer distance increases by about 0.3 \AA, indicating strong electrostatic repulsion.
While increasing the interlayer distance, the stacking energy is significantly lowered by 2.8 and 3.8 meV per atom for the two high-energy systems, respectively.

The results of bilayer and bulk are essentially comparable, with about twice the interaction energy found in the bulk compared to the bilayer due to the fact that each layer is interacting with two neighboring layers. 
The $c$ distances for the most stable AA' and AB stackings stay unchanged after optimization.

As next step, we have estimated the lowest-energy sliding path for the $h$-BN bilayer.
As this is a computationally demanding task, we have lowered the computational protocol in two aspects:
We do not perform the counterpoise correction, as we have noticed that the relative sliding energies are only slightly affected by BSSE for the BS3 basis.
Moreover, we noticed that the interlayer distance of the low-energy stacking orders do not change significantly, so we scan the potential energy surface at fixed interlayer distance of 3.31~\AA.
The result is given in Figure~\ref{fig:1} c. 
The resulting lowest-energy sliding path involves only one high-symmetry stacking, namely AA'; accessing the other low-energy stacking (AB) is only possible by rotation and subsequent translation of the basal planes with respect to each other.
Again, we have reoptimized the interlayer distance for several points along this path (Figure~\ref{fig:1} d), however, with only slight structural modification ($\le$0.06 \AA) and negligible energy change.
We estimate the minimum sliding energy to be 3.4 meV per atom. This value is almost 7 times higher than that reported by Marom et al.\cite{Marom2010}

Finally, we benchmark the performance of state-of-the-art DFT by comparing to our LMP2 reference calculations.
DFT overestimates the interlayer distance (as expected) due to missing London dispersion interactions.
An exception is found for the PBEsol functional that gives $c$ in a very good agreement with our LMP2 results and the experimental data (see Table~S1 in SI). The reason for this exceptional performance might be the inclusion of surface energies as target quantities when designing this functional.\cite{PBEsol}
Large $c$ results in the underestimation of the energy profile for all the functionals (see Figure~S3 in SI) except for PBEsol, that matches the LMP2 results with differences for the bulk (bilayer) of only 3.13 (1.25), 0.69 (0.56), and 1.57 (0.78) meV per atom for the high-energy AA, AB' and A'B structures, respectively (see Table~S2 in SI).
For the low-energy structures (AA' and AB), we find only negligible deviations between both methods (see Figures~\ref{fig:1} a and b).

Adding the correction for London dispersion (DFT-D calculations) strongly reduces $c$ (see Table~S1 in SI), shrinking it to values shorter than those found in the experiments and for LMP2.
Consequently, DFT-D is not suitable to treat the problem of stacking ordering in $h$-BN for the functionals studied in this work (see Figure~S4 in SI).
However, if the interlayer distance is correct, e.g.\ taken from a LMP2 calculation or experiment, DFT and DFT-D correctly describe the energy profile, with hybrid functionals being closer to the LMP2 reference values (see Figures~S5 and S6, and Table~S3 in SI).

In conclusion, we studied the stacking fault characteristics of the bulk and bilayer forms of hexagonal boron nitride by means of the local second-order M\o ller-Plesset perturbation theory (LMP2) for the solid state.
AA' is - in agreement with experiments - the lowest energy stacking order, with AB being slightly higher in energy. At thermally elevated temperature this energy difference will, however, be irrelevant. 
The AB' stacking is even less stable and thus unlikely to be observed in experiments on clean samples.
The lowest-energy sliding path for the bilayer passes only one high-symmetry stacking, namely the AA' equilibrium structure, and involves an energy barrier of 3.4 meV per atom.

The benchmark calculations with DFT(-D) methods show that the only functional that gives results comparable to our reference LMP2 calculations is PBEsol. Additional London dispersion interactions, however, ruin its performance and must not be applied for this functional in the context of $h$-BN studies.

The appearance of two local minima with little difference in relative energy explains the TEM results of Warner et al.,\cite{Warner2010} where islands of AB stacking were observed.
We finally remark that chemical interactions (Coulomb interaction and electron donation from N lone pairs to vacant B orbitals) are mainly responsible for the energy variation in the different stackings $h$-BN, while electron correlation softens the potential energy surface.

\providecommand*\mcitethebibliography{\thebibliography}
\csname @ifundefined\endcsname{endmcitethebibliography}
  {\let\endmcitethebibliography\endthebibliography}{}

\end{document}